\documentclass[showpacs,amsmath,amssymb,floatfix,aps,prl,reprint,groupedaddress,showkeys]{revtex4}
\usepackage{graphicx}
\usepackage{epsfig}
\usepackage{dcolumn}
\usepackage{bm}
\usepackage{times}

\begin{document}
\title{Single-electron Faraday generator}

\author{Gabriel Gonz\'alez}\email{gabrielglez@iteso.mx}

\affiliation{Departamento de Matem\'aticas y F\'isica, Instituto Tecnol\'ogico y de Estudios Superiores de Occidente, \\ Perif\'erico Sur Manuel G\'omez Mor{\'i}n 8585 C.P. 45604, Tlaquepaque, Jal., MEXICO}


\begin{abstract}
In this paper I study the posibility of inducing a single-electron current by rotating a non-magnetic conducting rod with a small tunnel junction immerse in a uniform magnetic field perpendicular to the plane of motion. I show first, by using a thermodynamic approach, the conditions needed to pump electrons around the mechanical device in the Coulomb blockade regime. I then use a density matrix approach to describe the dynamics of the single-charge transport including many-body effects. The theory shows that it is possible to have single-electron tunneling (SET) oscillations at low temperatures by satisfying conditions similar to the Coulomb blockade systems. 
\end{abstract}

\maketitle
\section{Introduction}
The field of single-electronics started when new effects due to the quantization of charge in ultrasmall tunnel junctions, both in the superconducting and the normal state, where predicted by Averin and Likharev \cite{lik1}. The theory of Averin and Likharev considers a tunnel junction which is biased by an externally fixed current $i$ and whose voltage $V$ is measured by a very high impedance voltmeter with metallic shunt conductance $G_S$. A tunnel junction consists of two conducting electrodes separated by a thin layer of insulating material and is characterized by its capacitance $C$ and tunnel resistance $R_T$. When a voltage is applied to the small capacitance tunnel junction, the charge will flow continuously through the conductor and it will accumulate on the surface of the electrode against the insulating layer of the junction (the adjacent electrode will have equal but opposite surface charge). On the other hand, the insulating layer is thin enough for electrons to tunnel through. The state of the junction is described by the surface charge $Q$ (which is a continuous variable) and the electrons $n$ that have tunnel through the insulating layer (which is a discrete variable). Averin and Likharev predicted that if the charge $Q$ at the junction is greater than $|e|/2$, an electron can tunnel through the junction in a particular direction, subtracting $|e|$ from $Q$. Likewise, if $Q$ is less than $-|e|/2$, an electron can tunnel through the junction in opposite direction, adding $|e|$ to $Q$. But if $Q$ is less than $|e|/2$ and greater than $-|e|/2$, tunneling in any direction would increase the energy of the system, hence tunneling will not occur. This suppression of tunneling is known today as the Coulomb blockade \cite{Hermann}. The physical origin of the Coulomb blockade of single-electron tunneling (SET) is quite simple. In a current-biased junction, each tunneling event leads to a change of the electrostatic energy of the system given by $\Delta E=e(Q \pm e/2)/C$. If the initial charge $Q$ is within the range $-|e|/2<Q<|e|/2$, the energy change $\Delta E$ is positive and at low temperatures tunneling events are impossible. On the other hand, if $|Q|>|e|/2$, tunneling is possible because this process reduces the electrostatic energy. An interesting prediction of Averin and Likharev was the SET oscillations in the voltage across the junction \cite{lik1, Hermann}. Due to the Coulomb blockade of tunneling the charge $Q$ on the junction accumulates until its threshold value $e/2$ is reached and then the junction is recharged by the externally fixed current. This whole process repeats itself with a frequency $\nu=i/e$ \cite{christoph}.\\
The purpose of this article is to show that there are SET oscillations without an external applied current in a small tunnel junction. In this case, the SET oscillations are driven by the Lorentz force due to the rotation of a conductor with a small tunnel junction and an applied external magnetic field. In addition, this mechanically driven device is proposed as a transducer of motion into electricity. 
The paper is organized as follows. First I will start by giving a thermodynamic formulation of the problem and the basic relations of the theory. Then I will analyze the system using a density matrix approach to include many-body effects and show the SET oscillations in the system. The conclusions are summarized in the last section.
\section{Thermodynamic Formulation}
\begin{figure}[ht]
  \begin{center}
    \begin{tabular}{cc}
      \resizebox{45mm}{!}{\includegraphics{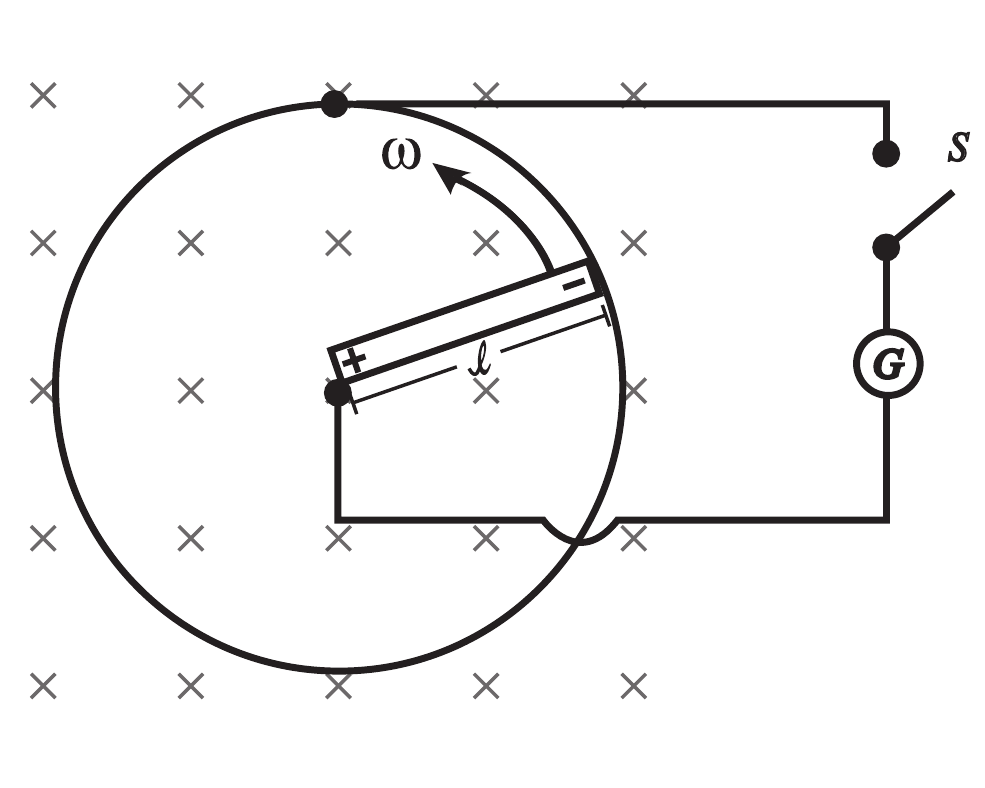}} &
      \resizebox{45mm}{!}{\includegraphics{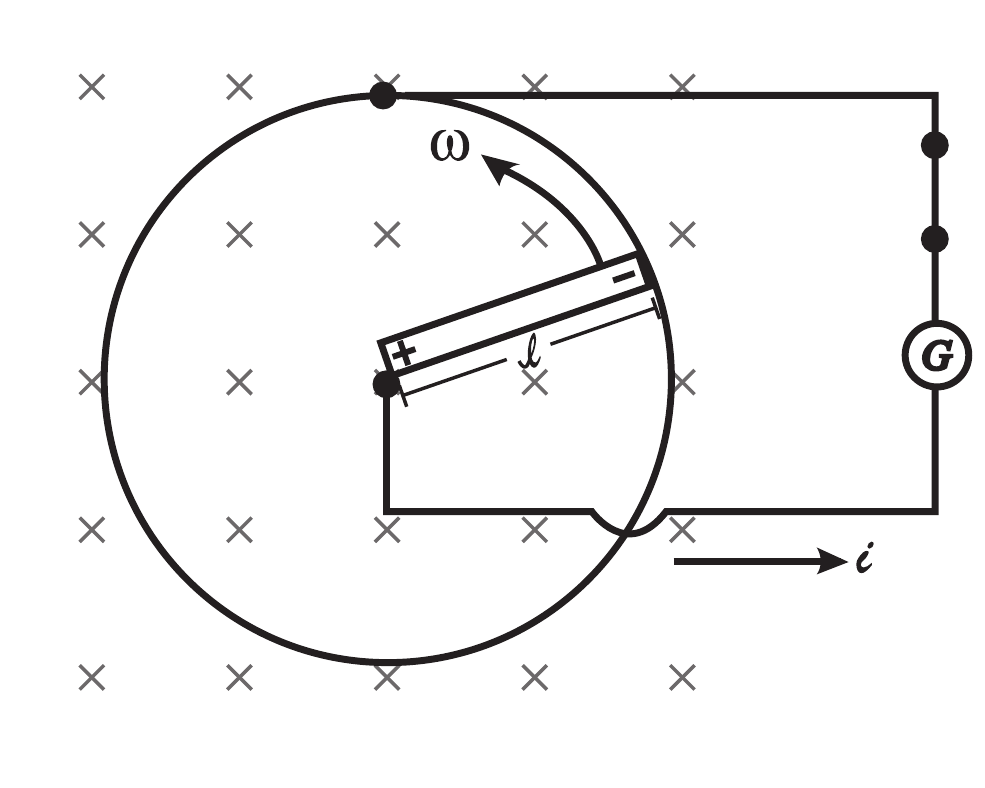}} \\ 
    \multicolumn{1}{c}{\mbox{\bf (a)}} &
		\multicolumn{1}{c}{\mbox{\bf (b)}} \\       
    \end{tabular}
\caption{Rotational motion of the conducting rod in the $XY$ plane
when the switch is (a) off or (b) on. The crosses
indicate that a uniform magnetic field is pointing into the page.}
\label{diagrams}
\end{center}
\end{figure}

\begin{figure}[!htb]
\begin{center}
  \includegraphics[width=7cm]{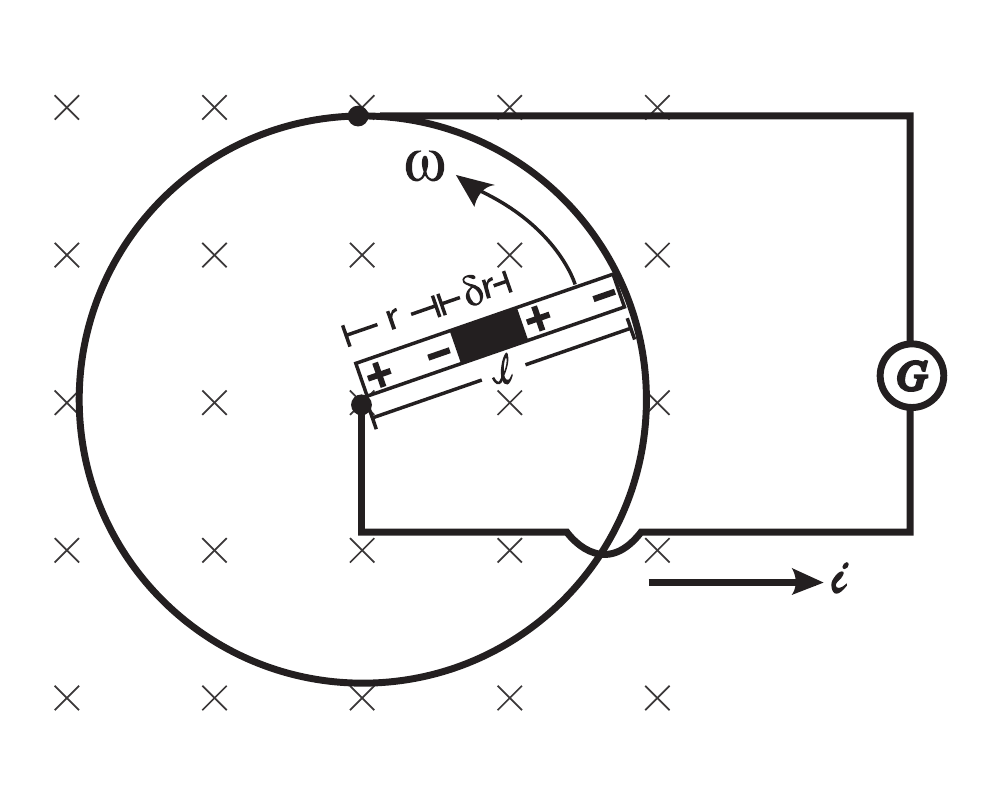}
  \caption[]{Schematic diagram for the rotation of the conducting rod with a tunnel junction of thickness $\delta r$ and a uniform magnetic field pointing into the page. Note how the charge accumulates on the surface of the electrode against the insulating layer. For this case current will only flow when a tunnel event occurs.} 
\label{tunnel}
\end{center}
\end{figure}

Consider the Faraday generator shown in Fig. (\ref{diagrams}), where a conducting rod of length $\ell$ rotates with constant angular velocity $\omega$ in a constant magnetic field that is perpendicular to the plane of motion. The rod completes the circuit, with one contact point on one end of the rod and the other on the circular rim. The circuit containing the galvanometer is completed by an open wire structure with a switch. 
For the case when the switch is open there will be no electromotive force, however we know from elementary electrodynamic courses that charge will pile up at the two ends of the rod and will produce an electric field that balances the Lorentz force felt by the moving charges inside the conductor (See Fig. (\ref{diagrams})(a)). \\
The free energy of the system is given by ${\cal F}=E-\vec{\mu}\cdot\vec{B}$, where $E$ is the total energy of the rotating rod, $\vec{\mu}$ is the magnetic dipole moment of the circulating charge at the end of the rod and $\vec{B}$ is the constant magnetic field perpendicular to the plane of motion. Taking the magnetic field as $\vec{B}=-B\hat{z}$, the free energy is given by 
\begin{equation}
{\cal F}=E_{in}+\frac{I\omega^2}{2}-\frac{QB\omega\ell^2}{2},
\label{eq1}
\end{equation}
where $E_{in}$ is the internal energy of the rod, $I$ denotes the moment of inertia for the rod with respect to the axis of rotation and we have taken the current as $i=Q\omega/2\pi$. It should be remembered that rotation in general changes the distribution of mass in the body, and so the moment of inertia and internal energy of the body are in general functions of $\omega$~\cite{lau}.\\
When the switch is turned on there is an electromotive force ${\cal E}$ in the circuit and hence charge will be circulating around the circuit. The free energy of the system in this case is given by
\begin{equation}
{\cal F}=E_{in}+\frac{I\omega^2}{2}-\frac{(Q-\Delta Q)B\omega\ell^2}{2}+{\cal E}\Delta Q,
\label{eq2}
\end{equation}
where $\Delta Q$ is the amount of charge circulating around the circuit and ${\cal E}\Delta Q$ is the work done by the system. The change in free energy is obtained by subtracting Eq.(\ref{eq2}) from Eq.(\ref{eq1}), which gives us 
\begin{equation}
\Delta {\cal F}=\left(\frac{B\omega\ell^2}{2}+{\cal E}\right)\Delta Q,
\label{eq3}
\end{equation}
if the system is in thermodynamic equilibrium then $\Delta {\cal F}=0$ and we obtain ${\cal E}=-B\omega\ell^2/2$, as we know from elementary electrodynamic courses \cite{ohanian}. Note that ${\cal E}<0$. \\
Now I will consider the case when there is a tunnel junction at position $r$ with thickness $\delta r$ in the conducting rod as shown in Fig.(\ref{tunnel}). 
For this case, even if the switch is turned on there will be no current flowing through the circuit because the charge $Q$ will accumulate on the surface of the electrode against the insulating layer as depicted in Fig.(\ref{tunnel}). Nevertheless, quantum mechanically speaking there is a probability for the charge to tunnel through the junction. The free energy of the system before quantum tunneling is given by ${\cal F}=E-\vec{\mu}\cdot\vec{B}-\vec{\mu}_1\cdot\vec{B}-\vec{\mu}_2\cdot\vec{B}$, where $\vec{\mu}$ and $\vec{\mu}_{1(2)}$ corresponds to the magnetic dipole moment of the charge accumulated at the end of the rod and against both sides of the insulating layer, and $E=I\omega^2/2+Q^2/2C$, where $C$ is the capacitance of the tunnel junction. Therefore, the free energy of the system before quantum tunneling is given by
\begin{equation}
{\cal F}=\frac{I\omega^2}{2}+\frac{Q^2}{2C}-\frac{QB\omega\ell^2}{2}-\frac{QB\omega r^2}{2}+\frac{QB\omega}{2}(r+\delta r)^2.
\label{eq5}
\end{equation}
Eq. (\ref{eq5}) can be written in the following form
\begin{equation}
{\cal F}=\frac{I^{\prime}}{2}\omega^2+\frac{Q^{\prime 2}}{2C},
\label{eq6}
\end{equation}
where
\begin{eqnarray}
I^{\prime}&=&I-C \left[\frac{B\ell^2}{2}-\frac{B\delta r(2r+\delta r)}{2}\right]^2 \\ \nonumber
Q^{\prime}&=&Q-C\omega\left[\frac{B\ell^2}{2}-\frac{B\delta r(2r+\delta r)}{2}\right]. 
\label{eq7}
\end{eqnarray}
When there is quantum tunneling there is a change in charge by $\pm|e|$ and an electromotive force ${\cal E}$ in the circuit which causes a change in the free energy given by
\begin{equation}
{\cal F}=\frac{I^{\prime}}{2}\omega^2+\frac{(Q^{\prime}\pm|e|)^2}{2C}+{\cal E}|e|.
\label{eq8}
\end{equation}
The change in free energy is obtained by subtracting Eq.(\ref{eq8}) from Eq.(\ref{eq6}), which gives us 
\begin{equation}
\Delta {\cal F}=\frac{|e|^2}{2C}\left(1\pm\frac{2Q^{\prime}}{|e|}\right)+|e|{\cal E}.
\label{eq9}
\end{equation}
Assuming ${\cal E}<0$ from our previous result, we see from equation (\ref{eq9}) that a tunnel event becomes energetically favorable and a current $i$ flows throughout the circuit when $Q^{\prime}>\frac{|e|}{2}$. When a tunnel event takes place the charge $Q^{\prime}$ will change by $-|e|$ and after a time $|e|/i$ the rotational motion of the conducting rod immerse in the constant magnetic field will recharged the junction and another tunnel event will take place. As a result the tunneling events will occur periodically with frequency $\nu=i/|e|$.\\
For the particular case in which the tunnel junction of thickness $\delta$ lies exactly in the middle of the conducting rod, i.e. $r=\ell/2-\delta/2$ and $r+\delta r=\ell/2+\delta/2$, then a tunnel event becomes energetically favorable when
\begin{equation}
Q>\frac{|e|}{2}+\frac{C\omega B\ell(\ell-\delta)}{2}.
\label{eq10}
\end{equation}
Eq. (\ref{eq10}) can be expressed in terms of the electrostatic voltage in the following way
\begin{equation}
V>\frac{|e|}{2C}+\frac{B\omega\ell(\ell-\delta)}{2}.
\label{eq11}
\end{equation}
Since the maximum voltage allowed for the system is given by $B\omega\ell^2/2$, then Eq. (\ref{eq11}) reads
\begin{equation}
\frac{B\omega\ell^2}{2}>V>\frac{|e|}{2C}+\frac{B\omega\ell(\ell-\delta)}{2}.
\label{eq11a}
\end{equation}
Eq. (\ref{eq11a}) gives us the following restriction $B\omega\ell\delta > |e|/C$. Using typical values of the tunnel junction capacitance $C\approx 3\times10^{-15}F$ and tunnel thickness $\delta=10\AA$ \cite{Hermann}, we need $B\omega\ell>10^{5}V/m$, to satisfy the restriction condition. If we have a magnetic field of $B=1T$ and $\ell=1cm$, then we will need a rotational frequency of around $\nu_r\geq10MHz$. The current delivered by the single-electron Faraday generator for this rotational frequency is around $i\approx 1pA$
\section{Many-body effects}
To study in more detail the dynamics of the charge transport we need the total Hamiltonian of the system depicted in Fig.~(\ref{tunnel}), which is given by:
\begin{equation}
{\cal H}={\cal F}(\hat{Q}^{\prime})+H_T+H_{1}+H_{2}+H_S-i\Phi.
\label{eqh}
\end{equation}
The first term in Eq. (\ref{eqh}) represents the free energy of the system which is given by Eq. (\ref{eq6}). The charge operator $\hat{Q}^{\prime}$ can be expressed via Fermion operators
\begin{equation}
\hat{Q}^{\prime}=-\frac{e}{2}\left(\sum_{k_1}c^{\dagger}_{k_1}c_{k_1}-\sum_{k_2}c^{\dagger}_{k_2}c_{k_2}\right)-Q_0,
\label{eqhf}
\end{equation}
where $c^{\dagger}_k$ and $c_k$ are the electron creation and annihilation operators and $Q_0=C\omega\left[\frac{B\ell^2}{2}-\frac{B\delta r(2r+\delta r)}{2}\right]$ is a constant term.
The second term in Eq. (\ref{eqh}) represents the tunneling Hamiltonian which is given by
\begin{equation}
H_T=\sum_{k_1,k_2}T_{k_1k_2}c^{\dagger}_{k_2}c_{k_1}+\sum_{k_1,k_2}T_{k_2k_1}c^{\dagger}_{k_1}c_{k_2},
\label{eqht}
\end{equation}
where the summation is carried out over all states $k$ within the electrodes 1 and 2 and $T_{k_1k_2}$ is the tunneling rate across the junction. 
The Hamiltonians $H_1$, $H_2$ and $H_S$ describe the energy of the internal degrees of freedom ${k_1}$, ${k_2}$ and ${k_S}$ of the two electrodes of the junction and of the shunt $G_S$, respectively. The last term in Eq. (\ref{eqh}) is the operator of the magnetic flux defined as
\begin{equation}
\Phi=-\int{\cal E}dt,
\label{eqm}
\end{equation}
where ${\cal E}$ is the electromotive force (emf) around the circuit. Note that Eq.(\ref{eqh}) corresponds exactly to the basic Hamiltonian given by Averin and Likharev. The only new feature is the shift of the charge operator $\hat{Q}^{\prime}=\hat{Q}-Q_0$ arising from the magnetic dipole interaction between the spinning charge and the external magnetic field. \\
Restricting ourselves to the case when the current through the junction and shunt are not too large we can consider them as perturbations and one can write an explicit time evolution equation for the density matrix to describe the junction properties. Following the pioneering work of Averin and Likharev \cite{lik1}, and assuming that $G_S, G_T<<4e^2/h$, the resulting master equation is given by 
\begin{equation}
\frac{\partial f}{\partial t}=F_T+F_S
\label{eqmaster}
\end{equation}
where $f(Q^{\prime},t)$ is the classical probability distribution and $F_T$ and $F_S$ are contributions due to the tunneling and shunt current, respectively, and are given by
\begin{eqnarray}
F_T(Q^{\prime})&=&\Gamma^{+}(Q^{\prime}-e)f(Q^{\prime}-e,t)+\Gamma^{-}(Q^{\prime}+e)f(Q^{\prime}+e,t) \nonumber \\ & & -[\Gamma^{+}(Q^{\prime})+\Gamma^{-}(Q^{\prime})]f(Q^{\prime},t)\\
F_S&=&\frac{G_S}{C}\frac{\partial}{\partial Q^{\prime}}\left(Ck_BT\frac{\partial f}{\partial Q^{\prime}}+fQ^{\prime}\right),
\label{eqf}
\end{eqnarray}
where $\Gamma^{\pm}$ are the tunneling rates for forward (plus sign) and backward (minus sign) single electron tunneling over the junction and can be expressed as
\begin{equation}
\Gamma^{\pm}(Q^{\prime})=\frac{1}{e}i(\Delta{\cal F}^{\pm}/e)\left[1-\exp\left(-\frac{\Delta{\cal F}^{\pm}}{k_BT}\right)\right]^{-1},
\label{eqt}
\end{equation}
where $i$ is the d.c. induced current and 
\begin{equation}
\Delta{\cal F}^{\pm}=\pm\frac{e}{C}\left(e/2\pm Q^{\prime}\right).
\label{eqF}
\end{equation}
Note that the master equation given in Eq. (\ref{eqmaster}) corresponds exactly to the master equation given in Averin and Likharev paper \cite{lik1}, the only difference is that there is no external current or bias.\\
If one looks at the regime $-e/2<Q^{\prime}<e/2$, the tunneling contributions may be neglected, i.e. $F_T=0$,  and Eq.(\ref{eqmaster}) can be expressed as
\begin{equation}
\frac{\partial f}{\partial t}=G_Sk_BT\frac{\partial^2}{\partial Q^{\prime 2}}\left[ f+\frac{V(t)}{k_BT}\right],
\label{eqdiff}
\end{equation}
where $V(t)$ is the time dependent voltage across the tunnel junction and is given by \cite{ferry}
\begin{equation}
V(t)=\frac{1}{C}\int Q^{\prime}f(Q^{\prime},t)dQ^{\prime}.
\label{volt}
\end{equation}
Making the substitution $F(Q^{\prime},t)=f(Q^{\prime},t)+V(t)/k_BT$, we end up with a reaction-diffusion equation given by
\begin{equation}
\frac{\partial F}{\partial t}=G_Sk_BT\frac{\partial^2 F}{\partial Q^{\prime 2}}+\frac{1}{k_BT}\frac{dV}{dt}.
\label{eqrd}
\end{equation}
For the system with constant electrostatic potential $V_0$, i.e. no tunneling, the solution to Eq. (\ref{eqrd}) is \cite{sneddon}
\begin{equation}
F=\frac{1}{\sqrt{t}}\exp[\frac{-Q^{\prime 2}}{4k_BTG_S t}]+\frac{V_0}{k_BT},
\label{eqrds}
\end{equation}
where the first term in Eq. (\ref{eqrds}) is the solution to the master equation for $f(Q^{\prime},t)$, which properly normalized can be expressed as
\begin{equation}
f(Q^{\prime},t)=\frac{1}{\sqrt{4\pi k_BTG_S t}}\exp[\frac{-Q^{\prime 2}}{4k_BTG_S t}].
\label{eqsol}
\end{equation}
Eq. (\ref{eqsol}) represents a Gaussian probability packet describing the distribution of charge $Q$ that will move due to the Lorentz force until $Q^{\prime}>e/2$, at this point the rate $\Gamma^{-}(Q^{\prime})$ becomes nonvanishing and this leads to a rapid decay of the packet, i.e. $f(Q^{\prime},t=t_T)=0$, where $t_T=C/G_T$ is the time when a tunneling event occurs. In this regime the tunneling event leads to a noticeable change in the voltage across the junction, i.e. $\Delta V=\pm e/C$, which is described in Eq. (\ref{eqrd}) by the last term, i.e.
\begin{equation}
\frac{\partial F}{\partial t}=G_Sk_BT\frac{\partial^2 F}{\partial Q^{\prime 2}}\pm\frac{e}{Ck_BT}\delta(t-t_T),
\label{eqrd1}
\end{equation}
The general solution to Eq. (\ref{eqrd1}) is just a shift of the solution given in Eq. (\ref{eqsol}) to the starting time $t=t_T$, i.e.
\begin{equation}
f(Q^{\prime},t)=\frac{\Theta(t-t_T)}{\sqrt{4\pi k_BTG_S (t-t_T)}}\exp[\frac{-Q^{\prime 2}}{4k_BTG_S (t-t_T)}],
\label{eqsol1}
\end{equation}
where $\Theta(t-t_T)$ is the Heaviside funtion. It is evident from Eq.(\ref{eqsol1}) that the whole process of the Gaussian packet formation repeats periodically with every tunneling event showing the periodic SET oscillations in the voltage across the junction. 
\section{Conclusions}
The main contribution of this article is to show that there can be SET oscillations across a tunnel junction without an externally applied current source. This result is in contrast to the system analyzed by Averin and Likharev where an external fixed current is always present. The thermodynamic and microscopic derivation shows how a single-electron current can be induced by rotating a conducting rod with a small tunnel junction in the presence of a uniform magnetic field perpendicular to the plane of motion. An estimate of the current delivered by the single-electron Faraday generator for rotational frequencies of $\nu_r\approx10MHz$ is $i\approx1 pA$. Thus, this device could serve as a fundamental standard of d.c. current.



\end{document}